\documentclass[a4paper,journal]{IEEEtran}

\usepackage{fix-cm} 
\usepackage[cmex10]{amsmath}
\usepackage{amssymb}
\usepackage{array}
\usepackage{booktabs}
\usepackage{multirow}
\usepackage{algorithm}
\usepackage{algorithmic}
\usepackage{cite}
\usepackage[dvips]{graphicx}
\usepackage{xfrac}
\usepackage{mathtools} 
\usepackage{cleveref}
\graphicspath{{figs/}}
\DeclareGraphicsExtensions{.eps}
\usepackage{color}
\usepackage[caption=false,font=footnotesize]{subfig}

\crefformat{equation}{(#2#1#3)}
\crefrangeformat{equation}{(#3#1#4) to~(#5#2#6)}
\crefmultiformat{equation}{(#2#1#3)}{ and~(#2#1#3)}{, (#2#1#3)}{ and~(#2#1#3)}

\newcommand{\hermit}{\mathrm{H}} 
\newcommand{\norm}[1]{\bigl\lVert#1\bigr\rVert}
\DeclarePairedDelimiter{\ceil}{\lceil}{\rceil} 

\begin{document}

\bstctlcite{IEEEexample:BSTcontrol}

\title{\LARGE  Projection-Based List Detection in\\Generalized Spatial~Modulation~MIMO~Systems}

\author{Jo\~ao~Cal-Braz,
        Raimundo~Sampaio-Neto
				
}

\maketitle


\begin{abstract}
This letter presents a novel detection strategy for Spatially-Multiplexed Generalized Spatial Modulation systems. It is a multi-stage detection that produces a list of candidates of the transmitted signal vector, sorted according to the proximity of the data vector to one of the possible vector subspaces. The quality metric and list-length metric selects the best candidate and manages the list length, respectively. Performance results show that it significantly reduces the performance gap to the optimal maximum likelihood detector, while maintaining significant computational cost reduction.

\end{abstract}

\begin{IEEEkeywords}
MIMO systems, Generalized Spatial Moduation, List Detection
\end{IEEEkeywords}

\section{Introduction}
Spatial modulation (SM) is a recently proposed multiple-antenna communication system that offers advantages over conventional MIMO systems, such as improved energy efficiency by means of the reduction of RF transmitting chains \cite{Renzo_commmag11}. It performs the transmission of information by means of the spatial dimension, coordinated by antenna indices, and the conventional signal constellation. Spatially Multiplexed Generalized Spatial Modulation (SMu-GSM) works as an extension of SM and aims to improve the spectral efficiency of SM \cite{Renzo13}. In this scheme, codewords are assigned to transmitting antenna combinations and each active antenna transmits independent symbols, as in a Spatial Multiplexing system.

Due to the infeasible implementation of the optimal detection scheme, several suboptimal approaches have been proposed \cite{Wang12,Xiao14,Cal-Braz14,Chen15}, however in all strategies a significant detection performance gap to the optimal detector is verified. An optimal detector for SMu-GSM, based on a sphere decoding (SD) algorithm has been proposed \cite{Cal-Braz14-2}, but the high computational cost in large scale systems or in low SNR regimes makes its implementation prohibitive in these scenarios.

List-based detectors have been extensively used in the detection of MIMO systems \cite{Waters08}. This scheme is often associated to conventional detection strategies, generating a list of symbol vectors candidates and a final decision is made using a decision metric. Several list-based approaches achieve near-ML performance~\cite{Bai09,Arevalo14,deLamare13}.


The work described herein proposes a novel suboptimal detection strategy for SMu-GSM systems. It consists of a multi-stage detection that sorts the signal vector candidates according the projection magnitudes on the vector subspaces related to the transmit antenna combinations. Specific forms of this strategy, based on the angle between the signal vector and the received data vector, have been developed for SM systems and are referred as signal vector based detectors~\cite{Wang12_2,Zheng12}. In the following, the generation of the candidates is conducted by a lattice-reduction zero-forcing (LR-ZF) detection, which is integrated to the candidates sorting method. A quality metric and a list-length metric compose the last stage. They are used to select the candidate as the detected transmitted vector and to update the list length, respectively. The aim of this detector is to provide better detection performance compared to existing strategies, reducing the gap to the optimal detector, while offering substantial computational complexity reduction.



\emph{Notation:} $\mathbf{A}^\hermit$ denotes the Hermitian transpose of a matrix $\mathbf{A}$, $\lfloor x \rfloor _{2^p}$ is the $p$-th integer power of two that is less than or equal to $x$. $\ceil{\cdot}$ represents the ceiling operation and $\{x_j\}_{j=1}^{N}$ denotes a set with elements $x_j$ indexed $j=1,2,\dotsc,N$.

\section {GSM System Model}
Consider a MIMO system with $N_T$ transmit antennas and $N_R$ receive antennas. In an SMu-GSM transmission, a subset containing $N_A$ of the transmit antennas are simultaneously activated to the emission of independent symbols from each antenna. So, in a channel use, $\log_2\left(N_C\right)$ bits are transmitted by the encoding of one of the $N_C$ antenna combinations into one codeword by consulting the spatial dimension mapping table. Additionally, $N_A$ symbols belonging to an alphabet of length $M$ are transmitted, totaling the transmission of $\log_2\left(N_C\right) + N_A\log_2\left(M\right)$ bits per channel use. The valid antenna combinations in the mapping table are indexed, $i = 1,2,\dotsc,N_C$, and $N_C = \lfloor \binom{N_T}{N_A} \rfloor_{2^p}$. The information symbols are transmitted through the $N_R \times N_T$ wireless channel $\mathbf{H}$. The received signal of this system is given by:

\begin{equation}
\label{sys_model}
\mathbf{y} = \mathbf{Hs} + \mathbf{n},
\end{equation}
where $\mathbf{s}$ is the $N_T \times 1$ data vector with zero entries and also elements belonging to a unit-energy discrete alphabet $\mathcal{S}$, and $\mathbf{n}$ is a zero-mean complex Gaussian vector with covariance matrix $\mathbb{E}\bigl[ \mathbf{nn}^\hermit \bigr] = \sigma^2 \mathbf{I}_{N_R}$. If the $k$th antenna combination is used for transmission, the equivalent $N_R\times N_A$ channel matrix $\mathbf{H}_k$ contains only the columns corresponding to the active antennas. Thus, \eqref{sys_model} can be reduced to:

\begin{equation}
\label{sys_model_red}
\mathbf{y} = \mathbf{H}_k\mathbf{x} + \mathbf{n},
\end{equation}
where $\mathbf{x}$ is the $N_A\times 1$ transmit vector, with all elements belonging to $\mathcal{S}$. Then, the average SNR per receive antenna is given by $\rho = \frac{N_A}{\sigma^2}$.



\section{Projection-Based List Detector (PBLD)}
The proposed strategy is structured in three steps that, altogether, compose a list detection algorithm. In the first stage, a scheme based on a projection filter bank sorts all valid antenna combinations in terms of the projections magnitudes. The following stage is devoted to the determination of symbol vector candidates, each assuming that one of the antenna combinations sorted in Stage~1 is correct. Finally, the last stage is composed of the selection of the best candidate calculated so far and a metric that defines the number of total candidates that should be generated in Stage~2 before terminating the algorithm. Perfect channel information at the receiver is considered. Further details of the detection stages and the execution flow are explained in the following subsections.

\subsection{Stage 1: Transmit Antenna Combination Sorting} \label{subsec:sorted_max_proj}

The strategy in this stage considers that the received vector tends to be closer to the space generated by the channel matrix used in the transmission, $\mathbf{H}_k$, whereas smaller projections are observed in the remaining $N_C-1$ subspaces spanned by $\mathbf{H}_i,\;i=1,2,\dotsc,N_C,\;i\neq k$. Then, the first stage is composed of a bank of $N_C$ linear filters. Each filter $\mathbf{W}_i$ is a matrix that orthogonally projects the input vector into the subspace spanned by $\mathbf{H}_i$, that is the channel matrix of one possible transmit antenna combination. Assuming $N_R \geq N_A$, the filter $\mathbf{W}_i, \, i = 1,2,\dotsc,N_C$ is given by:

\begin{equation}
\label{proj_mx_bank}
\mathbf{W}_i = \mathbf{H}_i\left(\mathbf{H}_i^\hermit\mathbf{H}_i\right)^{-1}\mathbf{H}^\hermit_i = \mathbf{H}_i\mathbf{G}_i^{\text{ZF}},
\end{equation}
where $\mathbf{G}_i^{\text{ZF}}$ is the zero-forcing (ZF) equalizer of a system with channel matrix $\mathbf{H}_i$. At the output of this bank, the transmit antenna combinations are sorted in descending order of the projection magnitude:

\begin{equation}
\label{max_proj_list}
\big\{ p_1, p_2,\dotsc,p_{N_C} \big\} = \arg\underset{i}{\operatorname{sort}} \norm{\mathbf{W}_i\mathbf{y}} = \arg\underset{i}{\operatorname{sort}} \norm{\mathbf{H}_i\mathbf{w}_i},
\end{equation}
where $\mathbf{w}_i = \mathbf{G}_i^{\text{ZF}}\mathbf{y}$ is an auxiliary vector.

\subsection{Stage 2: Symbol Vector Candidate Generation}
In this stage, a list of candidates of the transmitted symbol vector is generated. Nonetheless, the list length produced in Stage~2 is not fixed, as it is determined by the metric in Stage~3, to be presented. This incurs in the alternate execution of Stages 2 and 3.

An instance of the detector implemented in this stage is composed by a lattice-reduction (LR) operation, followed by a zero-forcing (ZF) equalization and vector detection. Then, for a candidate sorted as the $j$th in Stage~1, an $N_A\times N_A$ unimodular matrix $\mathbf{T}_{p_j}$ is computed by a basis reduction algorithm, that operates on the channel matrix $\mathbf{H}_{p_j}$ turning it into a new matrix $\tilde{\mathbf{H}}_{p_j} = \mathbf{H}_{p_j}\mathbf{T}_{p_j}$, with nearly orthogonal columns. This yields the effective received signal model \cite{Wubben04}:


\begin{equation}
\label{sys_model_lr}
\mathbf{y} = \tilde{\mathbf{H}}_{p_j}\mathbf{z}_{p_j} + \mathbf{n},
\end{equation}
where $\mathbf{z}_{p_j}$ is the effective symbol vector that results from the mapping $\mathbf{z}_{p_j} = \mathbf{T}^{-1}_{p_j}\mathbf{x}$. The equalization and quantization in LR domain can be considerably simplified considering that the product $\tilde{\mathbf{G}}_{p_j}^{\text{ZF}}\mathbf{y}$ can be written as:

\begin{align}
\label{lr_zf}
\tilde{\mathbf{G}}_{p_j}^{\text{ZF}}\mathbf{y}  &= \bigl( \tilde{\mathbf{H}}^\hermit_{p_j}\tilde{\mathbf{H}}_{p_j}\bigr)^{-1}\tilde{\mathbf{H}}^\hermit_{p_j}\mathbf{y} \nonumber \\
  &= \mathbf{T}^{-1}_{p_j}\bigl( \mathbf{H}^\hermit_{p_j}\mathbf{H}_{p_j}\bigr)^{-1}\mathbf{H}_{p_j}^\hermit\mathbf{y} \nonumber \\
	&= \mathbf{T}^{-1}_{p_j}\mathbf{G}_{p_j}^{\text{ZF}}\mathbf{y} = \mathbf{T}^{-1}_{p_j}\mathbf{w}_{p_j},
\end{align} 
where all possible vectors $\mathbf{w}_{p_j},\,j=1,\dotsc, N_C$, have already been calculated for application in \eqref{max_proj_list}. The quantization in LR domain after ZF equalization produces the detected vector:

\begin{equation}
\label{lr_quantize}
\hat{\mathbf{z}}_{p_j} = \mathcal{Q}_{\text{LR}} \bigl( \tilde{\mathbf{G}}_{p_j}^{\text{ZF}}\mathbf{y} \bigr) = \mathcal{Q}_{\text{LR}} \bigl( \mathbf{T}^{-1}_{p_j}\mathbf{w}_{p_j} \bigr),
\end{equation}
where $\mathcal{Q}_{\text{LR}}$ represents the quantization operation \cite{Milford11}. Then, the symbol vector candidate in the $\mathcal{S}^{N_A}$ space is retrieved applying the unimodular transformation:




\begin{equation}
\label{lr_domain_transf}
\mathbf{c}_{p_j} = \mathbf{T}_{p_j}\hat{\mathbf{z}}_{p_j}.
\end{equation}

Note that the minimum mean squared error (MMSE) equalization could be used instead. However, the savings in the number of computations by using the ZF filter makes this approach clearly advantageous in exchange for negligible performance loss.

\subsection{Stage 3: Candidate Election and List Length Update} \label{subsec:stage3}
The quality of the symbol vector candidate generated in Stage~2 is evaluated by its Euclidean distance to the received vector. The distance to the $j$th candidate, $\epsilon_{p_j}$, is given by:

\begin{equation}
\label{eq:quality_meas}
\epsilon_{p_j} = \norm{\mathbf{y} - \mathbf{H}_{p_j}\mathbf{c}_{p_j}}.
\end{equation}

The candidate quality is used to decide the amount of candidates that should be considered to perform a detection. A similar approach has been presented in a massive MIMO scenario, for the determination of the number of iterations allowed in a Monte Carlo Markov Chain detector in stalled state \cite{Datta12}. The length of the candidate list may be expressed in terms of a cost function that represents the closeness of the solution. It uses the statistics of $\epsilon^2$ for the case when $k$ and $\mathbf{x}$ are detected error-free. In this case, the cost is $\| \mathbf{n} \|^2$, which is Chi-squared distributed with $2N_R$ degrees of freedom, mean $N_R\sigma^2$ and variance $N_R\sigma^4$. The quality metric, $\phi(p_i)$, is defined as the difference between $\epsilon_{p_j}^2$ and the mean of $\| \mathbf{n} \|^2$, scaled by the standard deviation:

\begin{equation}
\label{quality_metric}
\phi(p_j) = \frac{\epsilon_{p_j}^2 - N_R\sigma^2}{\sqrt{N_R}\sigma^2}.
\end{equation}

The list-length metric, $\lambda$, is the result of the choice of an increasing function of $\phi(p_j)$, presented as an exponential function with growth rate $l_1$ and a lower-bound hard limit, $l_{\text{min}}$:

\begin{equation}
\label{list_length}
\lambda\bigl( \phi(p_j) \bigr) = \ceil*{ \max \bigl( l_{\text{min}},\exp\bigl( l_1\phi(p_j) \bigr) \bigr)  }.
\end{equation}

The parameters $l_\text{min}$ and $l_1$ share a single role: to avoid unnecessary iterations. As signal-to-noise ratio (SNR) increases, less often a candidate badly ranked in Stage~1 will be elected in Stage~3. Defining $c_{\text{lo}}$ and $c_{\text{hi}}$ as the minimum number of candidates to be processed in low and high SNR regimes ($\rho_\text{lo}$ and $\rho_{\text{hi}}$, respectively), the minimum number of candidates to be processed at an arbitrary SNR, $\rho$, is:

\begin{equation}
l_{\text{min}} = \frac{c_\text{hi} - c_\text{lo}}{\rho_\text{hi}-\rho_\text{lo}}\rho + c_\text{lo},
\end{equation}
where $c_\text{hi}$ and $c_\text{lo}$ ($c_\text{hi} < c_\text{lo}$) are fractions of the total number of antenna combinations. The growth rate parameter is made as $l_1 = \frac{l_\mathrm{min}}{\sqrt{\rho}}$.

\begin{figure}[!t]
\centering
\includegraphics[width=1.00\linewidth]{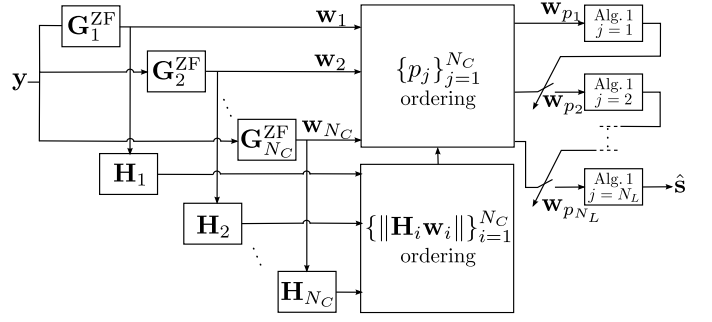}
\caption{Block diagram of the proposed strategy.}
\label{fig:blockdiag}
\end{figure}

Hence, when a new candidate, $j=J$, is obtained, this vector is considered the best current solution if the distance $\epsilon_{p_J}$ is smaller than the distances of the previous candidates, $j = 1,2,\dotsc,J-1$. In this case, the distance, $\epsilon_\text{min}$, of the best current solution, labeled as $\hat{\mathbf{x}}$, is updated. For this recently obtained candidate, the list length, $\Lambda$, is updated. The algorithm is terminated when the number of candidates generated equals $\Lambda$ \cite{Pereira14}. Supposing that the candidate $j=J$ was the best solution, the algorithm outputs the antenna combination candidate, $\hat{k} = p_J$, and the vector estimate, $\hat{\mathbf{x}}$. Then, the transmitted vector is reconstructed using the GSM mapping table, represented by $\mathcal{M}_\text{GSM}$:



\begin{equation}
\hat{\mathbf{s}} = \mathcal{M}_\text{GSM}\big( \hat{k},\hat{\mathbf{x}} \big).
\end{equation}


The sequential fashion the candidates are processed produces a never-increasing list, as it offers the possibility of list length reduction at every new best candidate found and, consequently, the reduction of the algorithm's computational cost. The block diagram of the strategy is presented in Fig.~\ref{fig:blockdiag} and the algorithm for implementing Stages 2 and 3 is presented in Alg.~1.



\algsetup{indent=1.35em}
\newcommand{\sdgesm}{\ensuremath{\mbox{\sc Detection algorithm: Stages 2 and 3}}}
\renewcommand{\algorithmicrequire}{\textbf{Input:}}
\renewcommand{\algorithmicensure}{\textbf{Output:}}
\begin{algorithm}[!t]
\small
\caption{$\sdgesm$}\label{gesm_sd_alg}
\begin{algorithmic}[1]
\REQUIRE $\mathbf{y}$, $\{p_j\}_{j=1}^{N_C}$, $\{\mathbf{w}_{p_j}\}_{j=1}^{N_C}$, $l_{\text{min}}$, $l_1$.
\ENSURE $\hat{k}$, $\hat{\mathbf{x}}$.
\medskip
\STATE $\epsilon_{\text{min}} \leftarrow \infty$ \quad $\Lambda \leftarrow N_C$ \quad $j \leftarrow 1$
\WHILE {$j \leq \Lambda$ \AND $j \leq N_C$}
	\STATE Generate candidate $\mathbf{c}_{p_j}$ using LR-ZF detection, using \cref{lr_quantize,lr_domain_transf}.
	\STATE Calculate distance to received vector, $\epsilon_{p_j}$, using \eqref{eq:quality_meas}.
	\IF {$\epsilon_{p_j} < \epsilon_{\text{min}}$}
			\STATE $\hat{\mathbf{x}} \leftarrow \mathbf{c}_{p_j}$ \quad $\hat{k} \leftarrow p_j$ \quad $\epsilon_{\text{min}} \leftarrow \epsilon_{p_j}$
			\STATE Calculate quality and list-length metrics, $\phi(p_j)$ and $\lambda\bigl( \phi(p_j) \bigr)$, using \eqref{quality_metric} and \eqref{list_length}.
			\STATE $\Lambda \leftarrow \lambda\bigl( \phi(p_j) \bigr)$
	\ENDIF
	\STATE $j \leftarrow j+1$
\ENDWHILE

\end{algorithmic}
\end{algorithm}

\section{Numerical Results}\label{sec:numresults}
In the simulation results, the channel matrix entries are random variables independently drawn from the zero-mean and unit-variance complex Gaussian distribution. The figures that follow are the result of $2\times 10^4$ simulation runs and 100 channel uses per run. In this section, performance and complexity comparisons of proposed strategy with existing schemes are presented. PBLD parameters were set as follows: $c_{\mathrm{lo}} = \frac{N_C}{4}$, $c_{\mathrm{hi}} = \frac{N_C}{8}$, $\rho_{\mathrm{lo}} = 0$~dB and $\rho_{\mathrm{hi}} = 30$~dB.

\begin{figure}[!t]
\centering
\includegraphics[width=1.00\linewidth]{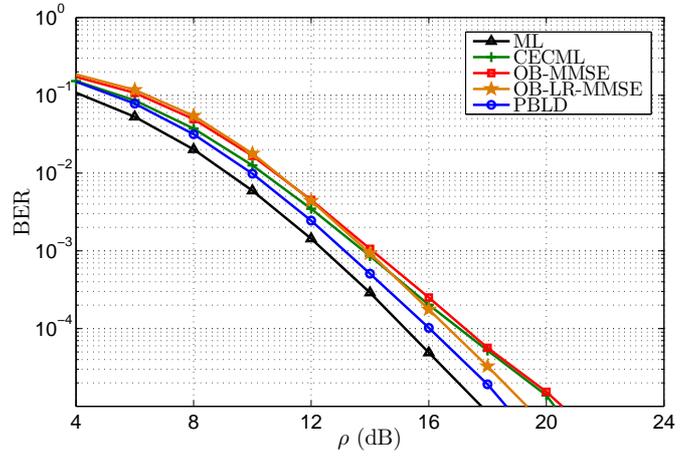}
\caption{Bit error rate for $N_T=4,\,N_A=2,\,N_R=4$,\,QPSK - 6~bits/chu.}
\label{teb1}
\end{figure}
\begin{figure}[!t]
\centering
\includegraphics[width=1.00\linewidth]{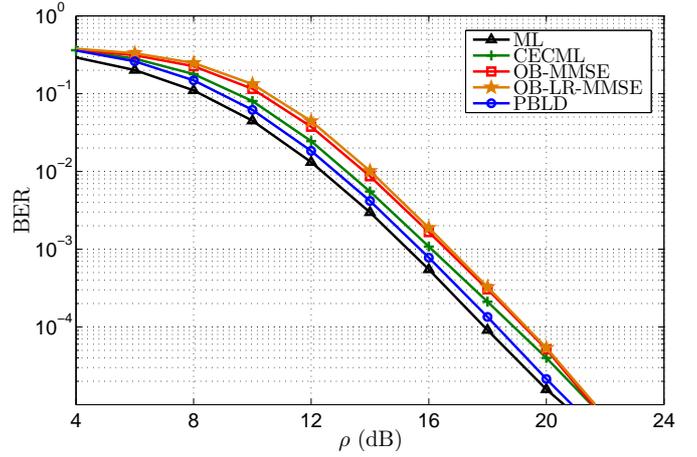}
\caption{Bit error rate for $N_T=16,\,N_A=2,\,N_R=4$, QPSK - 10~bits/chu.}
\label{teb2}
\end{figure}
\begin{figure}[!t]
\centering
\includegraphics[width=1.00\linewidth]{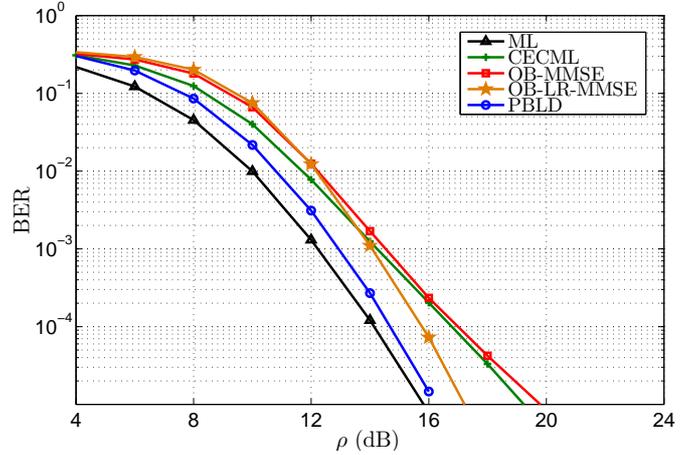}
\caption{Bit error rate for $N_T=7,\,N_A=4,\,N_R=7$, QPSK - 13~bits/chu.}
\label{teb3}
\end{figure}
\begin{figure}[!t]
\centering
\includegraphics[width=1.00\linewidth] {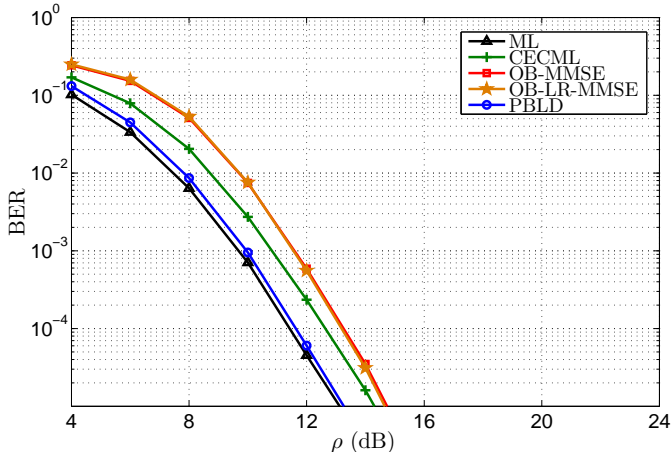}
\caption{Bit error rate for $N_T=8,\,N_A=3,\,N_R=8$, QPSK - 11~bits/chu.}
\label{teb4}
\end{figure}
\begin{figure}[!t]
\centering
\includegraphics[width=1.00\linewidth]{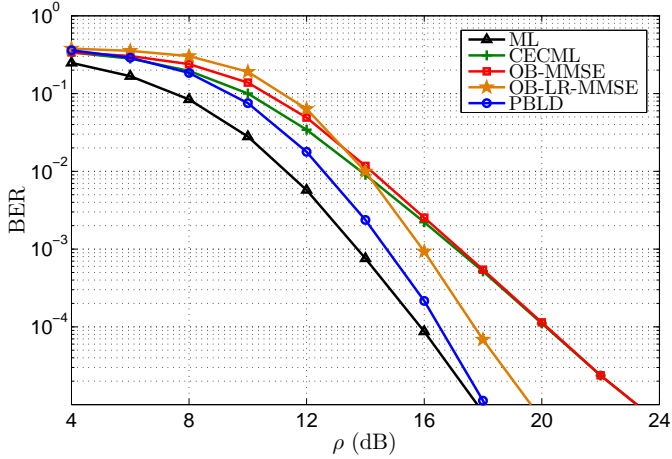}
\caption{Bit error rate under highly correlated channel ($\delta = 0.5$) for $N_T=7,\,N_A=4,\,N_R=7$, QPSK - 13~bits/chu.}
\label{teb3_corr0.5}
\end{figure}

In Figs.~\ref{teb1} to \ref{teb4}, the detection performance, in terms of the bit-error rate, of the proposed strategy is compared to OB-MMSE~\cite{Xiao14}, to an improved version of it, CECML~\cite{Chen15}, and to the strategy here labeled as OB-LR-MMSE. In the latter, we introduced lattice reduction processing in combination with the OB-MMSE strategy to produce the candidates. In all scenarios, the proposed scheme presents better detection performance than the listed alternatives. In particular, greatest advantages are observed in scenarios with larger number of antenna combinations and active transmit antennas, such as in Figs.~\ref{teb3} and \ref{teb4}, in which system configurations used were $N_T=7,\,N_A=4,\,N_R=7,$ QPSK, 13~bits/chu and $N_T=8,\,N_A=3,\,N_R=8,$ QPSK, 11~bits/chu, respectively. The behavior of the detection strategies under correlated channel was investigated and the result is shown in Fig.~\ref{teb3_corr0.5}. Using Kronecker channel correlation model~\cite{Kermoal02}, channel realizations are generated using $\mathbf{H}_{\mathrm{corr}} = \mathbf{R}_{\mathrm{RX}}^{1/2}\mathbf{H}\mathbf{R}_{\mathrm{TX}}^{1/2}$. The $(i,j)$th element of the correlation matrix, $\mathbf{R}_{\mathrm{TX}}$ or $\mathbf{R}_{\mathrm{RX}}$, is given by $r_{i,j} = \delta^{|i-j|^2}$, where $\delta$ is the correlation index between neighboring transmit or receive antennas. In this scenario, the advantage of the proposed strategy is maintained.


\begin{figure}[!t]
\centering
\includegraphics[width=1.00\linewidth]{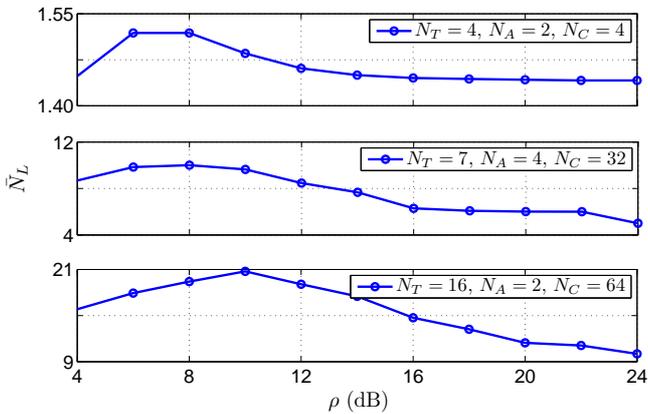}
\caption{Average list length $\bar{N}_L$.}
\label{listlength}
\end{figure}

\begin{figure}[!t]
\centering
\includegraphics[width=1.00\linewidth]{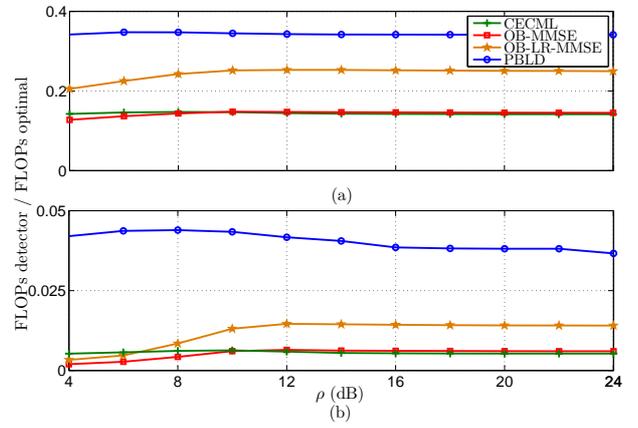}
\caption{Flop reduction for system configurations: (a) $N_T=4,\,N_A=2,\,N_R=4$,\,QPSK - 6~bits/chu, (b) $N_T=7,\,N_A=4,\,N_R=7$,\,QPSK - 13~bits/chu}
\label{flopreduc_varchan}
\end{figure}

The effect of the list length update scheme introduced in Sec.~\ref{subsec:stage3} is presented in Fig.~\ref{listlength}. The reduction of the average list length, $\bar{N}_L$, as the SNR increases is observed in the presented scenarios. Fig~\ref{flopreduc_varchan} shows the ratio between the number of floating point operations (FLOPs) required by the strategies and the optimal ML detector, evaluated in low and high rate system configurations (Figs. (a) and (b), respectively). Compared to the alternative schemes, PBLD achieves better performance at the cost of some computational complexity increase. Nonetheless, major savings in computational cost compared to the optimal ML detector are achieved.




\section{Conclusion}

A detection strategy for spatially multiplexed generalized spatial modulation MIMO systems was proposed. The scheme was shown to significantly reduce the gap to the optimal ML detector, while offering major computational cost savings compared to the optimal strategy.

\bibliographystyle{IEEEtran}
\bibliography{IEEEabrv,nested_ml_gesm_v2}

\end{document}